\documentstyle[vsolj02,graphicx,natbib]{article}

\RequirePackage[T1]{fontenc}

\def\cite{\citealt}
\setcitestyle{aysep={}}

\begin{document}

\title{Study of the low-amplitude Z~Cam star IX~Vel}

\author{Taichi Kato$^1$}
\author{$^1$ Department of Astronomy, Kyoto University,
       Sakyo-ku, Kyoto 606-8502, Japan}
\email{tkato@kusastro.kyoto-u.ac.jp}

\begin{abstract}
IX~Vel has been considered as one of the prototypical
novalike cataclysmic variables with thermally stable
accretion disks.  Using All-Sky Automated Survey (ASAS-3)
and All-Sky Automated Survey for Supernovae (ASAS-SN)
observations, I found that IX~Vel is a low-amplitude
dwarf nova showing standstills.  This object has been
re-classified as a Z~Cam-type dwarf nova.
This conclusion is consistent with the mass-transfer 
rate using the Gaia parallax which places the object
near the lower limit of the thermal stability.
Using two-dimensional Least Absolute Shrinkage and
Selection Operator (Lasso), I found that the cycle
lengths of dwarf nova outbursts varied between
13 and 20~d.  Analysis of the ASAS-3 data suggested
that the cycle lengths are shorter and the amplitudes
are smaller when the system is bright.
Standstills occurred when the system was bright.
These results support the idea that a subtle variation in
the mass-transfer rate from the secondary causes
transitions between outbursting state and standstills
in Z~Cam stars.
\end{abstract}

\section{Introduction}

   Despite its apparent brightness ($V \sim$9.5),
IX~Vel (=CD$-$48$^\circ$3636 = CPD$-$48$^\circ$1577) has been
relatively new to the field of cataclysmic variables (CVs)
[for general information of cataclysmic variables
and dwarf novae, see e.g. \citet{war95book}].
This object was spectroscopically identified as a UX UMa-type
novalike CV by \citet{gar84ixvel}.  \citet{egg84ixvel} performed
radial-velocity observations and obtained a possible
orbital period (0.1220~d, which was not a correct value).
\citet{war83ixvel} obtained a better orbital period
of 0.187(3)~d.  Although \citet{war83ixvel} reported
the similarity of the spectrum of IX~Vel with that of
the Z~Cam star HL~CMa, the lack of outbursts in the archival
plates suggested a novalike variable.  This reference was
the first to classify IX~Vel by long-term photometric history.
The modern binary parameters (orbital period = 0.193927~d)
were determined by \citet{beu90ixvel,kub99ixvel,lin07ixvel}.

   Using the data in All-Sky Automated Survey (ASAS-3: \cite{ASAS3}),
I found in 2018 that IX~Vel is a low-amplitude Z~Cam star
(vsnet-chat 8199).\footnote{
  $<$http://ooruri.kusastro.kyoto-u.ac.jp/mailarchive/vsnet-chat/8199$>$.
}  In this paper, I report analysis of IX~Vel as a dwarf nova.

\section{Results}

\subsection{Overall light curve}

\begin{figure*}
\begin{center}
\includegraphics[angle=0,width=16cm]{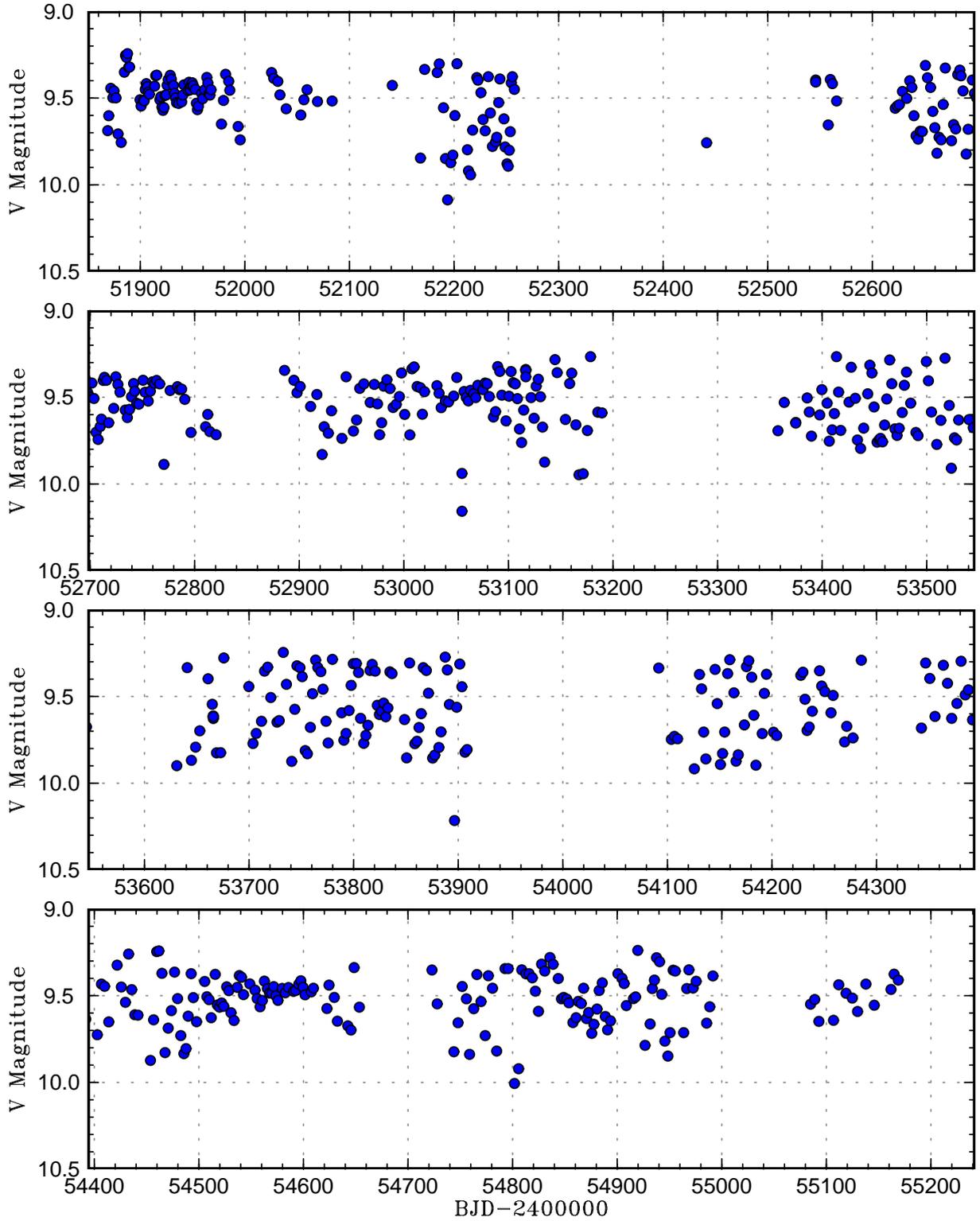}
\caption{
Light curve of IX~Vel using the ASAS-3 database.
}
\label{fig:ixvellc}
\end{center}
\end{figure*}

\begin{figure*}
\begin{center}
\includegraphics[angle=0,width=16cm]{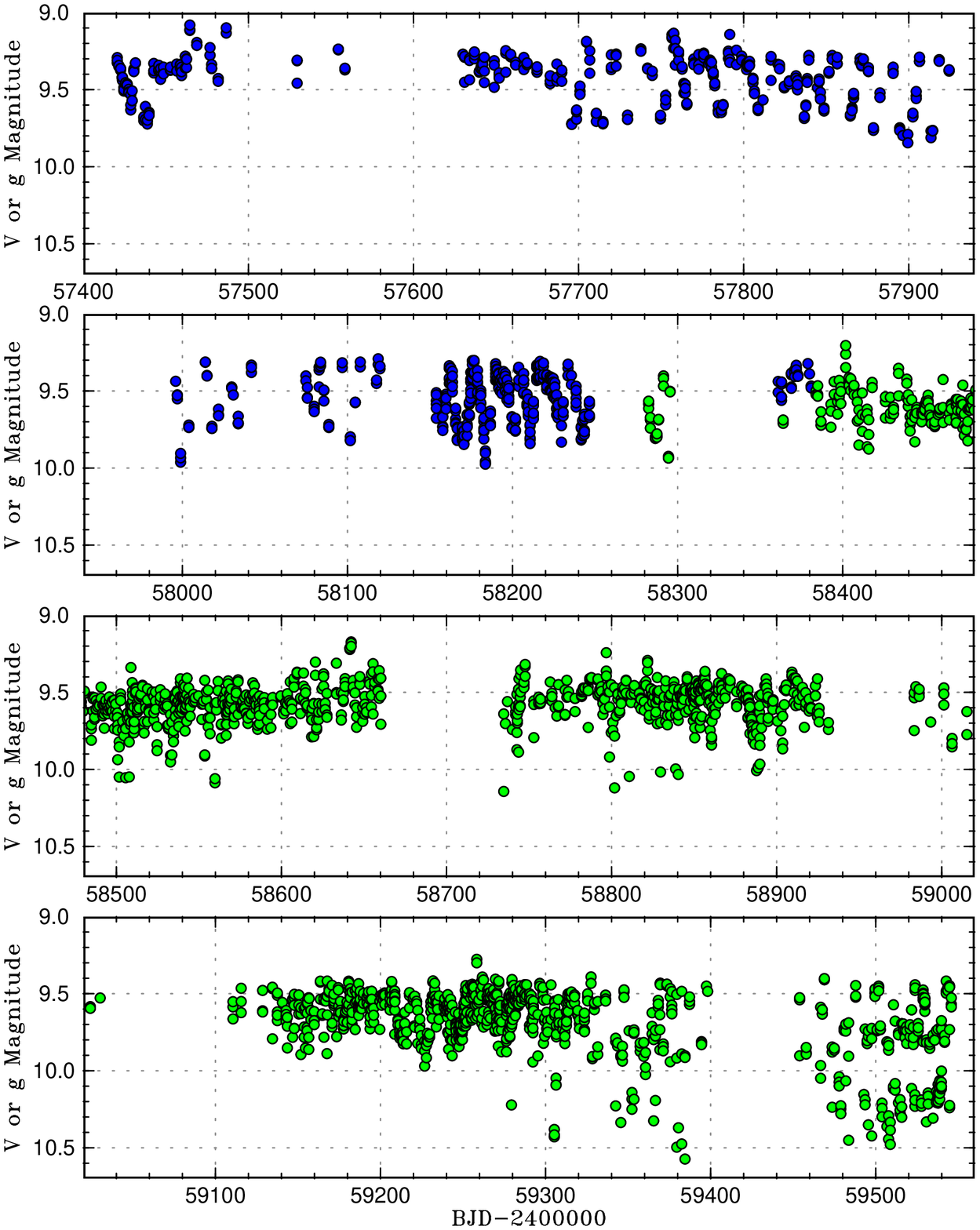}
\caption{
Light curve of IX~Vel using the ASAS-SN database.
Blue and green symbols represent $V$ and $g$ observations,
respectively.
Note that the scales are different from figure \ref{fig:ixvellc}.
}
\label{fig:ixvellc2}
\end{center}
\end{figure*}

   The light curve based on the ASAS-3 data is
shown in figure \ref{fig:ixvellc}.
Most of the time, this object showed low-amplitude
(up to 0.5~mag) quasi-cyclic variations.
The object, however, showed standstills such as
BJD 2453030--2453080 and 2454500--2454620.
The light curve based on the All-Sky Automated Survey
for Supernovae (ASAS-SN: \cite{ASASSN,koc17ASASSNLC})
is shown in figure \ref{fig:ixvellc2}.
Although some $g$-band observations of the ASAS-SN data
were apparently affected by saturation (particularly
after BJD 2459300 when there were a number of measurements
below 10.0 mag), the overall variations are composed of
phases of low-amplitude, quasi-cyclic variations and
standstills as in the ASAS-3 data.  The presence of
standstills are more apparent in the ASAS-SN data
(such as BJD 2457630--2457690 and BJD 2458440--2458630;
the amplitudes in more recent data were also small,
but this may have been affected by saturation).
The overall features are sufficient to classify IX~Vel
as a Z~Cam star.

\subsection{Analysis of dwarf nova-type outbursts using
   two-dimensional Lasso}

\begin{figure*}
\begin{center}
\includegraphics[angle=0,width=16cm]{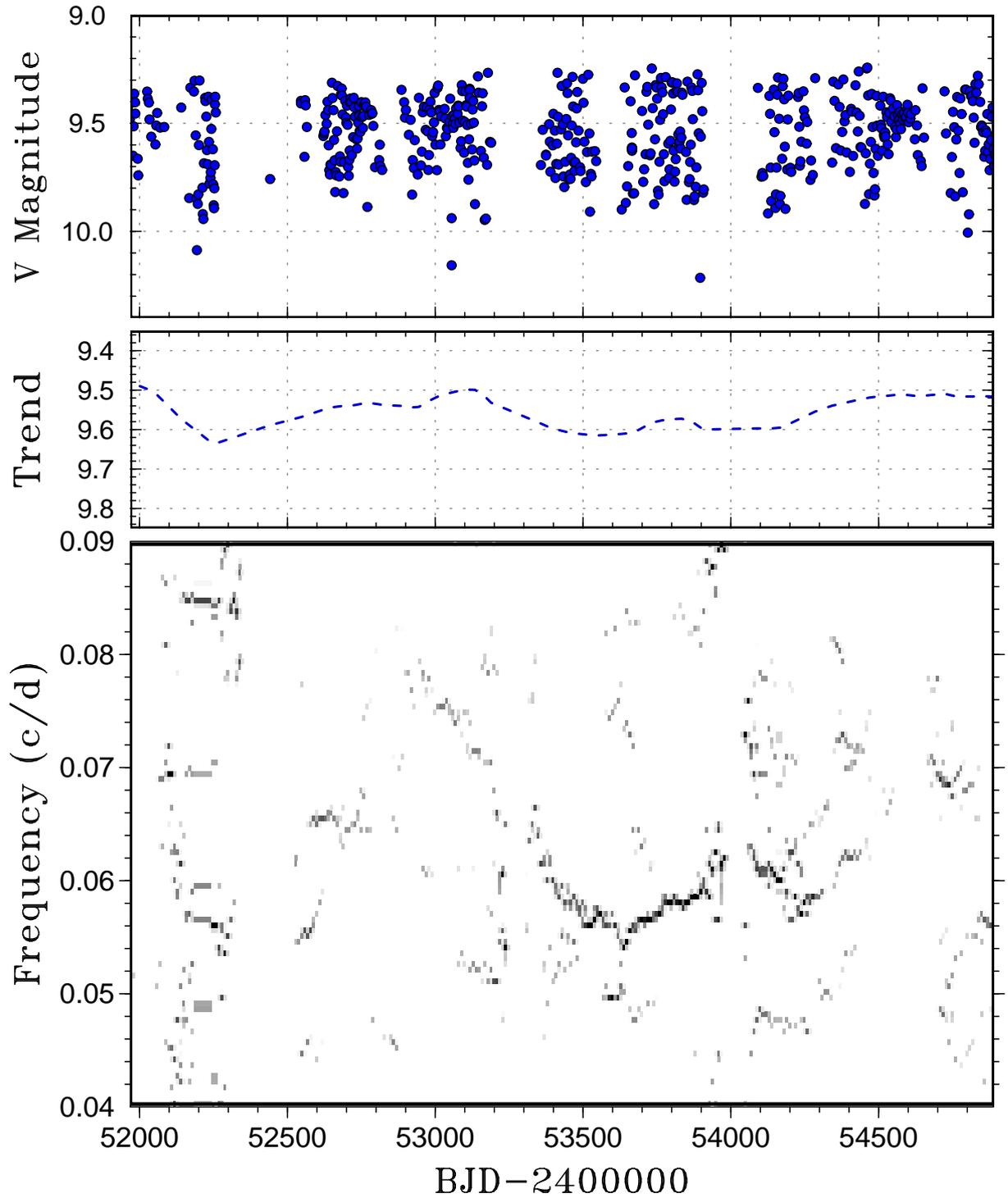}
\caption{
Period analysis of IX~Vel using the ASAS-3 database.
(Upper) Light curve.
(Middle) Trend of the light curve using LOWESS.
(Lower) Two-dimensional Lasso power spectrum.
The window size and shift value are 200~d and 10~d,
respectively.
}
\label{fig:ixvelspec}
\end{center}
\end{figure*}

   I used two-dimensional Least Absolute Shrinkage and
Selection Operator (Lasso) method
\citep{kat12perlasso,kat21lasso2} to study the variation
of cycle lengths in the dwarf nova phases.
The result for the ASAS-3 data is shown in
figure \ref{fig:ixvelspec}.  I used the window size and
shift value of 200~d and 10~d, respectively.
I had to use a relatively large window size due to
the limited sampling rate of observations.
During the interval BJD 2453400--2454300,
dwarf nova-type variations were most apparent at
frequencies 0.056--0.062~c/d, corresponding to
cycle lengths of 16--18~d.
The cycle lengths of dwarf nova-type variations
varied fairly strongly, and they were around
frequencies 0.071--0.078~c/d, corresponding to
cycle lengths of 13--14~d in the interval
BJD 2453000--2453200.  During the latter interval,
the signal in the two-dimensional Lasso spectrum
was weaker compared to the former interval, which can be
also seen as smaller amplitudes in the upper panel
of the figure.

   The middle panel of the figure shows the trend
of the light curve obtained by locally-weighted polynomial
regression (LOWESS: \cite{LOWESS}).
This demonstrates a positive correlation
between the brightness trend and the frequency of
the dwarf nova-type variations.  When the system was bright,
there was a tendency that the cycle lengths were shorter
and the amplitudes were smaller.  It is also apparent
from the figure that standstills occurred when
the system was bright.  It was a slight pity that
the two-dimensional Lasso spectrum was rather fragmentary
due to the presence of observational gaps.  It looks likely
that the cycle lengths would have been longer when
the system was faint around BJD 2452300, but the lack of
the data prevented me from drawing a conclusion.

\begin{figure*}
\begin{center}
\includegraphics[angle=0,width=16cm]{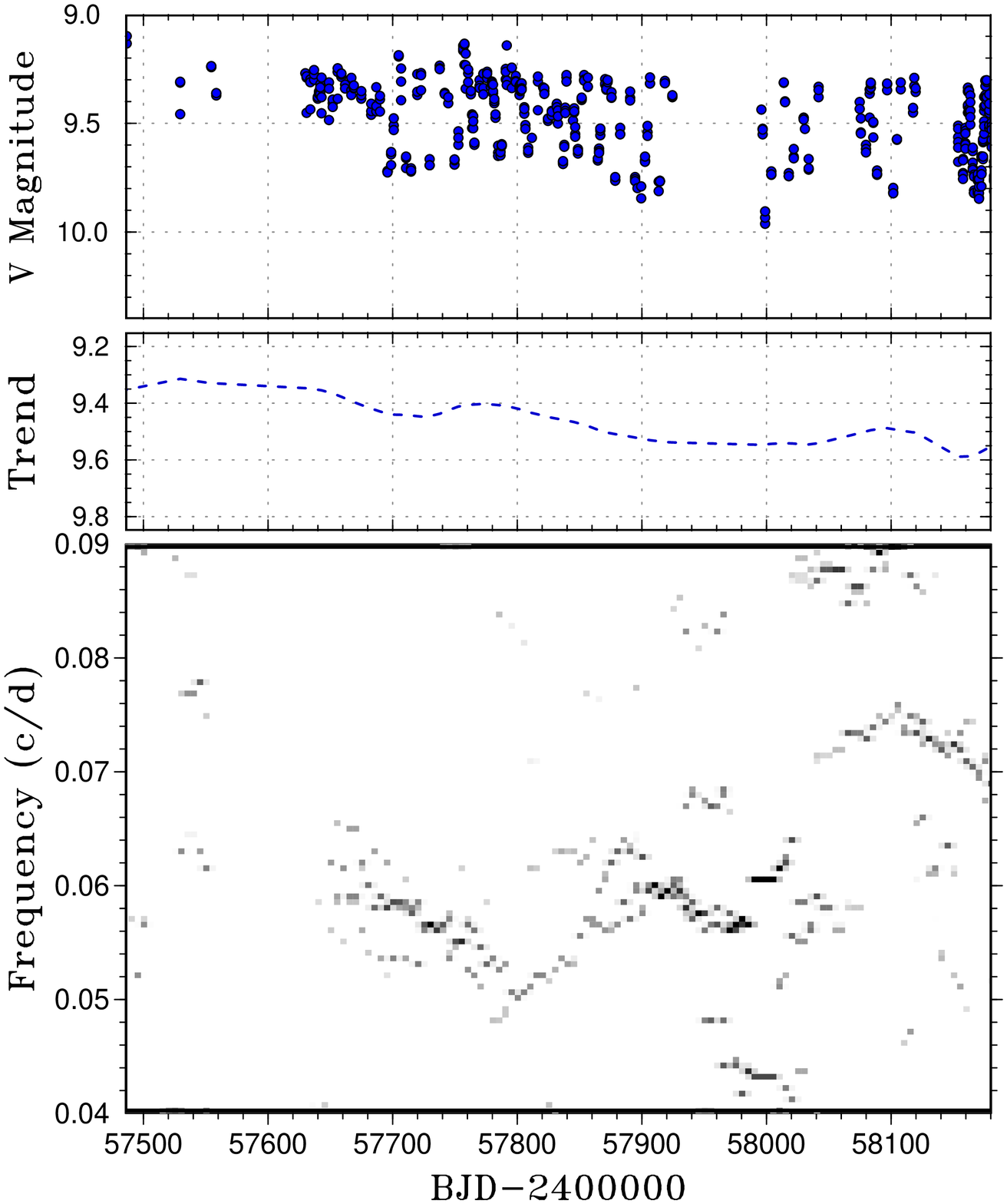}
\caption{
Period analysis of IX~Vel using the ASAS-SN $V$ data.
(Upper) Light curve.
(Middle) Trend of the light curve using LOWESS.
(Lower) Two-dimensional Lasso power spectrum.
The window size and shift value are 130~d and 5~d,
respectively.
}
\label{fig:ixvelspec2}
\end{center}
\end{figure*}

   The same type of analysis for the ASAS-SN data is shown
in figure \ref{fig:ixvelspec2}.  In order to avoid
the effect of saturation, I used only $V$ data.
Readers should keep in mind that the entire time interval
of the figure is much shorter than the ASAS-3 data
(figure \ref{fig:ixvelspec}).  Thanks to the higher
sampling rate of the ASAS-SN observations,
I could use a smaller window size and shift value
of 130~d and 5~d, respectively.

   In the upper panel of the figure, the dwarf nova-type
variations are clearly seen after BJD 2457700.
The frequencies varied between 0.05~c/d (corresponding to
a cycle length of 20~d) to 0.075~c/d (13~d).
In contrast to the ASAS-3 data, this variation of
the frequencies does not look like to positively correlate
with the trend of the light curve.  This result is based
on a relatively short segment of data (700~d, compared
to 2900~d in the ASAS-3 data) and needs to be treated
with caution.  I consider that the result from the ASAS-3
data is more reliable in discussing the correlation.

\section{Discussion}

   The mechanism of transition between outbursting state and
standstills in Z~Cam stars is not still well understood.
There has been a consensus that a subtle variation
in the mass-transfer rate from the secondary causes
transitions between outbursting state and
standstills \citep{mey83zcam}.
This mechanism implies that the mean brightness during
standstills is brighter than during outbursting state.
In four of five ``classical'' Z~Cam stars studied by
\citet{hon98zcam}, the standstills were as bright as
or brighter (by 0.2 mag in some systems)
than the mean brightness during outbursting states.
The unusual standstills of SY~Cnc were
reported to be fainter than the mean brightness during
outbursting intervals \citep{hon98zcam}.  Whether
standstills in Z~Cam stars are universally brighter
than outbursting states still needs to be investigated.
As I have shown, the ASAS-3 data of IX~Vel indeed show this
tendency and this would be an additional support to
the idea by \citet{mey83zcam}.

   In the case of IX~Vel, the mass-transfer rate appears
to be close to the limit of thermal instability and
when the mass-transfer rate is slightly below
the critical value, dwarf nova-type outbursts occur.
The mass-transfer rate, however, is only slightly below
the critical value and only the outer part of the disk
is thermally unstable (or the cooling wave cannot easily
reach deeper into the inner region of the disk).
This explains why the amplitudes of dwarf nova-type
outbursts in IX~Vel are so small.
According to \citet{deb18CVgaia},
the mass-transfer rate of IX~Vel using Gaia DR2
data \citep{GaiaDR2}\footnote{
   The result is unchanged using Gaia EDR3 \citep{GaiaEDR3}.
} is near the lower limit of
the thermal stability and the above interpretation is
consistent with the disk-instability theory.

   Low-amplitude outbursts during standstills of
``classical'' Z~Cam stars have also been documented,
such as in \citet{szk84AAVSO,kat01zcam}.
These phenomena are likely the same as low-amplitude
outbursts recorded in IX~Vel.

   \citet{kim20iwandmodel} performed 
numerical simulations of dwarf nova outbursts
(both in tilted and non-tilted disks) based on
the thermal-viscous instability model
(the simulation scheme was based on \cite{ich92diskradius}).
In their simulations (figure 18 in \cite{kim20iwandmodel}),
tilted disks are prone to thermal instability even
under the condition of the mass-transfer rate high enough
to keep the entire disk hot in a non-tilted disk.
Their resultant light curve looks similar to that of
IX~Vel and it would be worth studying whether there
is a signature of a disk tilt in this system.

   The present observational knowledge of the behavior of
the accretion disk near the border of the thermal instability
might also be helpful in understanding the unusual behavior
of IW And stars
\citep{sim11zcamcamp1,ham14zcam,kat19iwandtype,kim20iwandmodel},
which are also considered to have accretion disks near
the border of the thermal instability.

\section*{Acknowledgements}

This work was supported by JSPS KAKENHI Grant Number 21K03616.
I am grateful to ASAS-3 and ASAS-SN teams for
for making the database available to the public.

\newcommand{\noop}[1]{}\newcommand{\hyphalt}{-}


\begin{thebibliography}{}

\bibitem[{Beuermann} and {Thomas}(1990)]{beu90ixvel}
  {Beuermann}, K., \& {Thomas}, H.-C. (1990) Detection of emission lines from
  the secondary star in {IX Velorum} (={CPD}$-$48$^\circ$1577). {\em A\&A\,}
  {\bf 230}, 326

\bibitem[{Cleveland}(1979)]{LOWESS}
  {Cleveland}, W.~S. (1979) {Robust locally weighted regression and smoothing
  scatterplots}. {\em J. Amer. Statist. Assoc.\,} {\bf 74}, 829

\bibitem[{Dubus} et~al.(2018)]{deb18CVgaia}
  {Dubus}, G., {Otulakowska-Hypka}, M., \& {Lasota}, J.-P. (2018) Testing the
  disk instability model of cataclysmic variables. {\em A\&A\,} {\bf 617}, A26

\bibitem[{Eggen} and {Niemela}(1984)]{egg84ixvel}
  {Eggen}, O.~J., \& {Niemela}, V.~S. (1984) {CoD}$-$48$^\circ$3636 : an
  apparently bright, low-luminosity and high-temperature variable. {\em AJ\,}
  {\bf 89}, 389

\bibitem[{Gaia Collaboration} et~al.(2018)]{GaiaDR2}
  {Gaia Collaboration} {et~al.} (2018) {Gaia Data Release 2}. {Summary} of the
  contents and survey properties. {\em A\&A\,} {\bf 616}, A1

\bibitem[{Gaia Collaboration} et~al.(2021)]{GaiaEDR3}
  {Gaia Collaboration} {et~al.} (2021) {Gaia Early Data Release} 3. {Summary}
  of the contents and survey properties. {\em A\&A\,} {\bf 649}, A1

\bibitem[{Garrison} et~al.(1984)]{gar84ixvel}
  {Garrison}, R.~F., {Schild}, R.~E., {Hiltner}, W.~A., \& {Krzeminski}, W.
  (1984) {CPD}$-$48$^\circ$1577: the brightest known cataclysmic variable. {\em
  ApJ\,} {\bf 276}, L13

\bibitem[{Hameury} and {Lasota}(2014)]{ham14zcam}
  {Hameury}, J.-M., \& {Lasota}, J.-P. (2014) Anomalous {Z Cam} stars: a
  response to mass-transfer outbursts. {\em A\&A\,} {\bf 569}, A48

\bibitem[Honeycutt et~al.(1998)]{hon98zcam}
  Honeycutt, R.~K., Robertson, J.~W., Turner, G.~W., \& Mattei, J.~A. (1998)
  Are {Z Camelopardalis}-type dwarf novae brighter at standstill?. {\em PASP\,}
  {\bf 110}, 676

\bibitem[Ichikawa and Osaki(1992)]{ich92diskradius}
  Ichikawa, S., \& Osaki, Y. (1992) Time evolution of the accretion disk radius
  in a dwarf nova. {\em PASJ\,} {\bf 44}, 15

\bibitem[Kato(2001)]{kat01zcam}
  Kato, T. (2001) {\noop{Kato2001ibvs5093}} {Oscillation} during a standstill
  of {Z Cam}. {\em IBVS\,} {\bf 5093}, 1

\bibitem[{Kato}(2019)]{kat19iwandtype}
  {Kato}, T. (2019) Three {Z Cam}-type dwarf novae exhibiting {IW And}-type
  phenomenon. {\em PASJ\,} {\bf 71}, 20

\bibitem[{Kato}(2021)]{kat21lasso2}
  {Kato}, T. (2021) {\noop{Kato2021vsolj86}}{A} code for two-dimensional
  frequency analysis using the {Least Absolute Shrinkage} {and Selection
  Operator} ({Lasso}) for multidisciplinary use. {\em VSOLJ\ Variable\ Star\
  Bull.\,} {\bf 86}, (arXiv:2111.10931)

\bibitem[{Kato} and {Uemura}(2012)]{kat12perlasso}
  {Kato}, T., \& {Uemura}, M. (2012) Period analysis using the {Least Absolute
  Shrinkage and Selection Operator} ({Lasso}). {\em PASJ\,} {\bf 64}, 122

\bibitem[{Kimura} et~al.(2020)]{kim20iwandmodel}
  {Kimura}, M., {Osaki}, Y., {Kato}, T., \& {Mineshige}, S. (2020)
  Thermal-viscous instability in tilted accretion disks: toward understanding
  {IW And}-type dwarf novae. {\em PASJ\,} {\bf 72}, 22

\bibitem[{Kochanek} et~al.(2017)]{koc17ASASSNLC}
  {Kochanek}, C.~S. {et~al.} (2017) {The All-Sky Automated Survey for
  Supernovae} ({ASAS-SN}) light curve server v1.0. {\em PASP\,} {\bf 129},
  104502

\bibitem[{Kubiak} et~al.(1999)]{kub99ixvel}
  {Kubiak}, M., {Pojmanski}, G., \& {Krzeminski}, W. (1999) Spectroscopic
  observations of {IX Vel}. {\em Acta\ Astron.\,} {\bf 49}, 73

\bibitem[{Linnell} et~al.(2007)]{lin07ixvel}
  {Linnell}, A.~P., {Godon}, P., {Hubeny}, I., {Sion}, E.~M., \& {Szkody}, P.
  (2007) A synthetic spectrum and light-curve analysis of the cataclysmic
  variable {IX Velorum}. {\em ApJ\,} {\bf 662}, 1204

\bibitem[Meyer and Meyer-Hofmeister(1983)]{mey83zcam}
  Meyer, F., \& Meyer-Hofmeister, E. (1983) A model for the standstill of the
  {Z Camelopardalis} variables. {\em A\&A\,} {\bf 121}, 29

\bibitem[{Pojma\'nski}(2002)]{ASAS3}
  {Pojma\'nski}, G. (2002) The {All Sky Automated Survey}. {Catalog} of
  variable stars. {I}. 0$^{\rm h}$--6$^{\rm h}$ quarter of the southern
  hemisphere. {\em Acta\ Astron.\,} {\bf 52}, 397

\bibitem[{Shappee} et~al.(2014)]{ASASSN}
  {Shappee}, B.~J. {et~al.} (2014) The man behind the curtain: {X}-rays drive
  the {UV} through {NIR} variability in the 2013 {AGN} outburst in {NGC 2617}.
  {\em ApJ\,} {\bf 788}, 48

\bibitem[{Simonsen}(2011)]{sim11zcamcamp1}
  {Simonsen}, M. (2011) The {Z CamPaign}: Year 1. {\em J.\ American\ Assoc.\
  Variable\ Star\ Obs.\,} {\bf 39}, 66

\bibitem[Szkody and Mattei(1984)]{szk84AAVSO}
  Szkody, P., \& Mattei, J.~A. (1984) Analysis of the {AAVSO} light curves of
  21 dwarf novae. {\em PASP\,} {\bf 96}, 988

\bibitem[{Wargau} et~al.(1983)]{war83ixvel}
  {Wargau}, W., {Drechsel}, H., {Rahe}, J., \& {Bruch}, A. (1983)
  Spectrophotometry of the recently discovered cataclysmic variable
  {CPD}$-$48$^\circ$1577. {\em MNRAS\,} {\bf 204}, 35P

\bibitem[Warner(1995)]{war95book}
  Warner, B. (1995) Cataclysmic Variable Stars (Cambridge: Cambridge University
  Press)

\end{thebibliography}
\end{document}